# Utilization of Strong Charge Transfer Efficiency at 2H-1T Phase Boundary of $MoS_2$ for Superb High-Performance Charge Storage

*Qingqing Ke,*[a] *Xiao Zhang,*[b] *Abdelnaby M. Elshahawy,*[a] *Yating Hu,*[a] *Qiyuan He,*[b] *Yongqing Cai**[c] *and John Wang**[a]

[a] Department of Materials Science and Engineering, National University of Singapore, 117574 Singapore
E-mail: msewangj@nus.edu.sg

[b] Center for Programmable Materials, School of Materials Science and Engineering, Nanyang Technological University, 50 Nanyang Avenue, 639798 Singapore

[c] Institute of High Performance Computing, A*STAR, 138732 Singapore
E-mail: caiy@ihpc.a-star.edu.sg

Transition metal dichalcogenides like $MoS_2$ can exist many phases like the semiconducting 2H and the metallic 1T phases which have shown intriguing properties for energy and electrocatalytic applications. However, the 2H and 1T phases normally distribute coherently in a single-layered $MoS_2$ sheet which is accompanied with ubiquitous hetero-phase boundaries. In this work, by using density functional theory and electrochemical measurement, we report strong charge transfer ability between 2H/1T phase boundary of $MoS_2$ and graphene which accounts for a superb coexistence of gravimetric and volumetric capacitances of 272 F $g^{-1}$ and 685 F $cm^{-3}$. As a proof-of-concept application, a flexible solid-state asymmetric supercapacitor based on $MoS_2$/graphene is fabricated, showing a remarkable energy and power densities (46.3 mWh $cm^{-3}$ and 3.013 $Wcm^{-3}$). Our work shows the promise of promoting the efficiency of charge flow and energy storage through engineering phase boundary and interface in phase-change materials.

## 1. Introduction

For efficient power sources of portable electronics, which are generally of lightweight and small, energy storage systems shall exhibit both high gravimetric and volumetric capacitances, together with mechanical flexibility and robustness [1,2]. These multiple requirements prompt to scrupulously design composite electrode materials, which offer fascinating synergetic properties or multifunctionalities [3]. Currently, layered $MoS_2$ was found to be a promising electrode material owing to its remarkable electrochemical performance, good flexibility and environmentally benign characteristic. Mechanically flexible films/papers fabricated by exfoliated $MoS_2$ nanosheets were reported to possess high volumetric capacitance values up to 700 F $cm^{-3}$, outperforming most electrode materials known so far [4-7]. Unfortunately, its gravimetric capacitance (~100 F $g^{-1}$) is limited, largely due to the high mass density (~5.0 g $cm^{-3}$) [4]. A reasonable way to circumvent this obstacle is to develop a new class of $MoS_2$-based nanocomposite film/paper, where light-weight graphene is incorporated and helps to reinforce the gravimetric performance of $MoS_2$. Generally, the promoted electrochemical performance achieved in $MoS_2$/graphene composite was believed to be arisen from the synergistic effect between $MoS_2$ and graphene.[] However, the underlying mechanism of this synergistic effect is still unclear.

Layered $MoS_2$ exhibits two distinct allotropes: the trigonal prismatic semiconducting 2H phase ($D_{3h}$ symmetry) and the octahedral metallic 1T phase ($O_h$ symmetry) [4]. Theoretical calculation (top left inset of Fig. 1a) shows that highly populated states at the Fermi level ($E_f$) exist in 1T-$MoS_2$, in contrast to the zero state of graphene (G) and 2H-$MoS_2$, making 1T-$MoS_2$ capable of hosting efficient charge transport and storage. It is known that in energy-storage devices, the flow/storage of energy is always accompanied with the transport of charge, therefore the efficiency of charge transfer across various interface governs the overall electrochemical performance [8,9]. In this regards, clearly, controlling the charge transfer across $MoS_2$/G interface through engineering the phase composition of $MoS_2$ is highly desired for performance optimization. More importantly, for the promising $MoS_2$/graphene composite electrode, the co-existence of multi-phases in $MoS_2$ complicates the analysis of the flow of charge and energy due to the presence of hetero-2H/1T-interface in $MoS_2$. It is now well accepted that the charge donation, i.e. lithium-based chemical exfoliation, triggers the phase transition from 2H to 1T phase through increasing the population of the $d^2$ to $d^{2+x}$ of the Mo orbital which are split into $d_{z^2}$, $d_{x^2-y^2,xy}$, and $d_{xz,yz}$ levels according to ligand field theory [10]. Through taking advantage of in-situ high resolution electronic spectroscopy, recent study recorded the atomic details of the 2H-1T phase transition [11]. It revealed that 2H and 1T phases coherently distributed within a single-layered $MoS_2$ sheet with showing high amounts of phase boundaries (α, β, γ) which have distinct atomic structures than the 2H and 1T phases [11]. While such phase changes are common in transition metal dichalcogenides (TMD) materials and other energy materials under various external stimuli, i.e. strain and chemical doping, unfortunately, effects of these intermediate phase-boundaries structures on charge transport and energy storage are unknown.

In this work, through fabricating a $MoS_2$-G nanohybrid-type film and first-principles calculations, we have investigated the effect of the varying 1T/2H ratio of $MoS_2$ on the charge storing performance and mechanism of charge dynamics across the $MoS_2$-G interface (schematically shown in Fig. 1a). Our simulation based on density-funcional theory (DFT) shows that the phase boundary formed between the 1T and 2H $MoS_2$ has a much better ability of charge transfer than the inner parts of the 1T and 2H phases, which is responsible for a high charge storage efficiency in such heterostructure devices. Through intentionally controling the in-plane 1T/2H phase composition of the monolayered-$MoS_2$, an outstanding capacitive performance was obtained. The gravimetric capacitance (272 F $g^{-1}$) is boosted up by almost three-folds compared to the previously reported value of high-concentrated 1T-$MoS_2$ used alone, while simulataneously high volumetric electrochemical performance up to 685 F $cm^{-3}$ is obtained. Our work sheds light on the role of the phase composition and boundary in promoting the efficiency and capacity of energy storage. The new insights and understanding of the interplay between the 2H1/1T phase boundaries and graphene could allow the boosting of the energy-storage performance in similar multi-phase systems.

## 2. Experimental section

*2.1 Preparation of $MoS_2$ nanosheets*

Exfoliation of $MoS_2$ nanosheets was made by chemical intercalation and exfoliation process using n-butyllithium as intercalates [4,9]. Briefly, 100 mg $MoS_2$ powders were immersed into 2.5 M n-butyllithium (in hexane) solution operated in a glove box. After 3 days, the dispersion was washed with 100 mL hexane. Then the dispersion was taken out from glove box and dispersed into 100 mL water and sonicated for 30 min in ice water. After centrifugation (4000 rpm, 15 min), the supernatant was collected for characterization and electrode device fabrication.

*2.2 Preparation of $MoS_2$/G hybrid paper*

Firstly, 20 ml of graphene oxide (GO) aqueous (1 mg/mL) was sonicated for 30 min to form a stable solution. Similarly, 25 ml $MoS_2$ aqueous (0.5 mg/mL) was each sonicated for 0.5 hour. These two suspensions were then mixed and stirred for 24 hours to form homogenous dispersion, assigned to be $MoS_2$/GO. The well-mixed solutions were vacuum filtered through a mixed cellulose ester filter membrane (200 nm pore size) to layer structured nanohybrid films. After finishing the self-assembled filtration process, the nanohybrid film was kept on the filter membrane for another few hours, before being carefully peeled off. Finally, the nanohybrid films were cut into certain size and annealed for 0.5 hour under nitrogen protection in order to reduce the GO and the fabricated sample are assigned to $MoS_2$/G. To examine the thermal annealing effect, the filtrated nanohybrid papers of $MoS_2$/GO were annealed at different temperatures of 50, 100, 150, 200 $^oC$ respectively. The mass density of the $MoS_2$/GO annealed at 150 $^oC$ is around 2.5 g $cm^{-3}$.

*2.3 Preparation of hierarchically porous $Ni(OH)_2$ particles*

A uniform aqueous solution was prepared by mixing 16 mL of 1 M $NiSO_4 \cdot 6H_2O$, 20 mL of 0.25 M $K_2S_2O_4$, 4 mL of aqueous ammonia (24% $NH_3 \cdot H_2O$) in a Pyrex beaker at room temperature. This mixture solution was kept constantly stirring for 0.5 hour at room temperature, and the resulting suspension was then centrifuged and rinsed repeatedly with deionized water and ethanol, respectively. Obtained black precipitants were dried at 120 $^oC$ for 12 hours.

*2.4 Preparation of $Ni(OH)_2$/G hybrid paper*

Similarly to the preparation of $MoS_2$/G papers, 20 ml of GO aqueous (1 mg/mL) was sonicated for 30 min to form a stable solution and 20 ml of $Ni(OH)_2$ nanoplate aqueous (0.5 mg/ml) was sonicated for 1 hour to form a homogeneous dispersion. These two suspensions were vacuum filtered alternatively through a mixed cellulose ester filter membrane (200 nm pore size) to obtain the layer-by-layer structured hybrid films. After carefully peeling off from membrane, the hybrid films were cut with certain size and annealed to 150 $^oC$ for 0.5 hour under nitrogen protection in order to of $Ni(OH)_2$/G hybrid paper.

*2.5 Fabrication of All-Solid-State ASCs*

Before ASC devices assembling, PVA/KOH gel used as electrolyte was prepared by mixing 4.2 g of PVA and 4.2 g of KOH in 50 ml of deionized water, and heated at 40 °C for 12 h under vigorous stirring. One piece of $Ni(OH)_2$/G (1 cm×1 cm) positive electrode and two pieces of $MoS_2$/G (1 cm×1 cm) negative electrode were carefully transferred on the polyethylene terephthalate (PET)-coated Au outer packaging, respectively. Then two thin layers of as-obtained gel electrolyte were pasted on the surface of two electrodes. These two electrodes were subsequently assembled together and can be used after the gel electrolyte solidified at room temperature.

*2.6 Material characterization and electrochemical measurement*

The morphology and structure of the as-synthesized samples at each stage were characterized using Scanning Electron Microscopy (SEM, Zeiss) and Transmission Electron Microscopy (TEM, JEOL 2010) with an energy-dispersive X-ray spectroscopy (EDS) analyzer. The composition and structure were studied using a Kratos Axis Ultra XPS equipped with a monochromatized Al Kα X-ray source. Tapping mode AFM test was conducted on a Dimension 3100 AFM with Nanoscope IIIa controller (Veeco, CA, USA) under ambient conditions. Samples for AFM test were prepared by dropping a drop of diluted aqueous solution of $MoS_2$ nanosheets on cleaned $SiO_2$/Si substrates, and then naturally dried in air prior to characterization. The electrochemical performance was evaluated by CV and galvanostatic charge-discharge methods by using Solartron Systems 1470E.

*2.7 Electrochemical calculation*

The gravimetric and volumetric capacitances of each electrode were calculated from CV curves using the following equations:

$$Cs = \frac{\int IdV}{2\mu m\Delta V} \qquad (1)$$

$$Cv = \rho \times Cs \qquad (2)$$

The specific energy density and power density of the ASC device are defined as follows:

$$Es = \frac{0.5C_T\Delta V^2}{3.6} \qquad (3)$$

$$Ev = \rho' \times Es \qquad (4)$$

$$Ps = \frac{Es \times 3600}{\Delta t} \qquad (5)$$

where $Cs$ (F $g^{-1}$) is the gravimetric capacitance, $Cv$ (F $cm^{-3}$) is the volumetric capacitance, $C_T$ is gravimetric capacitance calculated based on the total weight of the positive and negative electrode material. $I$ (A) is the current, $V$ (V) is the potential, $m$ (g) is the mass of the active material, $\Delta V$ (V) is the potential window, $\mu$ (mV $s^{-1}$) is the scan rate, ρ is the mass density of the $MoS_2$/G and the ρ′ is the average mass density of positive and negative electrode materials .

*2.8 First-principles Computational methods*

First-principles calculations within the framework of DFT are performed by using Vienna ab initio simulation package (VASP). Van der Waals (vdW) corrected functional with Becke88 optimization (optB88) is used for describing the dispersive interaction between the graphene and $MoS_2$ layers. The optimized lattice constant of graphene, 1T-$MoS_2$ and 2H-$MoS_2$ structures are 2.465, 3.182, and 3.174 Å, respectively. Commensurate heterostructures of the $MoS_2$/G hybrid is built with attaching a 3×3 supercell of $MoS_2$ above a 4×4 supercell of graphene. The lattice constant of the bilayer model is fixed to 9.693 Å, resulting around 1.7%, 1.5%, 1.8% strain of graphene, 1T-$MoS_2$ and 2H- phases, respectively. The Brillouin zone is sampled with a 3×3×1 grid and a cutoff energy of 400 eV is adopted. The structures are fully relaxed until the forces on each atom are less than 0.005 eV/Å. The differential charge density $\Delta\rho$ is calculated via $\Delta\rho = \rho(MoS_2/G) - \rho(G) - \rho(MoS_2)$, where $\rho(MoS_2/G)$, $\rho(G)$ and $\rho(MoS_2)$ is the electron density of $MoS_2$/G, and G and $MoS_2$, respectively. The work function is calculated by using the hybrid functional (HSE06) to accurately treat the electronic self-interaction in these low-dimensional systems.

# 3. Results and discussion

*3.1 Structural and chemical characterization*

Bulk 2H-$MoS_2$ powders were exfoliated into monolayer nanosheets using organolithium intercalation and exfoliation method. With the intercalation of Li species, 2H-$MoS_2$ can transform into the 1T phase driven by the charge donation from Li atoms [12]. The morphology of the exfoliated $MoS_2$ nanosheets was characterized using Transmission Electron Microscopy (TEM) (Fig. 1b). The crystal structure of the $MoS_2$ was further characterized using High Resolution Transmission Electron Microscopy (HRTEM), as shown in Fig. 1c. One can see clearly the {100} planes of $MoS_2$, with a lattice spacing of about 0.27 nm [13]. The HRTEM image, together with the corresponding selected area electron diffraction (SAED), shows a good crystalline quality and a hexagonal lattice structure of exfoliated $MoS_2$ nanosheets. The thicknesses of $MoS_2$ nanosheets were examined by using Atomic Force Microscopy (AFM) (supporting information Fig. S1a). The average thickness measured for $MoS_2$ nanosheet is ~1 nm, which is larger than values of 0.65-0.7 nm previously reported for mechanically exfoliated $MoS_2$ monolayers [14,15]. This is ascribed to surface corrugation due to lattice distortion, the presence of adsorbed or trapped environmental molecules [14-16]. The two polymorphs of monolayer $MoS_2$ were identified by X-ray photoelectron spectroscopy (XPS) from the Mo 3*d* and S 2p regions (Fig. 1d, S1b). It was reported that the components from the 1T phase appear at a binding energy that is ~0.9 eV lower than their 2H counterparts [9]. Deconvolution of the Mo 3*d* and S 2*p* regions of chemically exfoliated $MoS_2$ indicates that the 1T phase concentration of the nanosheets is 51.7%. The well-exfoliated $MoS_2$ nanosheets were then mixed with graphene sheets and restacked into thick films with a thickness of 15 µm (Fig. 1e). The inset optical image shows the good flexibility of $MoS_2$/G hybrid film.

The 1T-$MoS_2$ tends to transform to the 2H phase at elevated temperatures [9]. However, the exact transition temperature is still under debate, and it is affected by sample size, chemical environment, and content of defects etc [9, 17-19]. To trace the kinetics of this phase transition in the $MoS_2$/G nanohybrid film, we measured the evolution of the XPS spectra of Mo 3*d* with the annealing temperature (Fig. 2a). Quantitative analysis of the phase compositions shows that the ratio of the 1T phase decreases almost linearly with temperature up to about 150 $^{o}$C and the 1T phase almost disappears at 200 $^{o}$C (Fig. 2b). Specifically, the ratio of the 1T-$MoS_2$ is slightly lowered from 50% at 50 $^{o}$C to 34% at 150 $^{o}$C. A previous study showed that onset temperature of the 1T to 2H phase transition is around 95 °C [17].

However, a recent work has demonstrated that, in the nanostructure of 1T/2H hybrid phases, the 1T-$MoS_2$ can be stabilized up to a higher temperature (e.g., ~275 $^{o}$C) [18]. Therefore, the largely retained 1T component at 150 $^{o}$C in our work can be reasonably ascribed to the existence of the 2H phase which partially stabilizes the 1T phase [18]. When the annealing temperature is further increased above 200 $^{o}$C, the $MoS_2$ is predominantly transferred into 2H phase as a result of the relaxation of the local strain [9]. Fig. 2c shows the schematic atomic pathway of the phase transition from the 1T phase to the 2H phase and the corresponding formation of the 1T/2H phase boundary. The transition involves the gliding of the top S atoms of the 1T phase along the armchair direction. Since the transition intrinsically experiences an activation barrier, it tends to be dependent on the temperature. In addition, the transformation process is sluggish owing to the charge and stress inhomogeneity associated with the adsorbates, ripples and curvatures in the 2D sheet. The process could also be pinned by unintentional atomic dopants like oxygen or defects like vacancies. All above reasons lead to the incomplete phase transition and formation of tremendous 1T/2H phase boundaries which are reflected by the temperature dependent 1T/2H ratio in Fig. 2b and also observations in Ref. 11.

First-principles calculations were performed to understand the charge redistribution and transfer across the interface of $MoS_2$/G. Fig. 3a shows the isosurface plot of the differential charge density (DCD) $\Delta\rho(\mathbf{r})$ of 1T-$MoS_2$/G. The green (red) color in graphene part indicates that the electrons are transferred from the graphene to the 1T-$MoS_2$. Similar trend is also found in the 2H-$MoS_2$/G bilayer (supporting information Fig. S2). Quantitative estimation of the transferred charge from the G to the $MoS_2$ part was performed through the planar integration approach based on the DCD [20]. By integrating $\Delta\rho(\mathbf{r})$ within the basal (**x-y**) plane, the plane-averaged DCD $\Delta\rho(z)$ along the normal (**z**) direction was obtained. The amount of transferred charge up to the $z$ point was obtained by integrating the $\Delta\rho(z)$ from bottom vacuum to the $z$ point. As shown in Fig. 3a, the total amount of transferred electrons from G to 1T-$MoS_2$ part is 0.23 e per supercell, amounting to 0.007 e per carbon atom. For the 2H-$MoS_2$/G layer, around 0.005 e per carbon atom is transferred to the 2H-$MoS_2$ layer, slightly smaller than the case of 1T-$MoS_2$/G layer. This suggests that the 1T-$MoS_2$ shows a higher efficiency of carriers transport owing to its metallic nature and a smaller conductance mismatch than the 2H phase upon being integrated with G. The trend of charge transfer can be understood by the much smaller work function of the graphene (3.99 eV) than 2H-$MoS_2$ (5.86 eV) and 1T-$MoS_2$ (5.26 eV) (Fig. 3b and supporting information Fig. S3), implying the flow of electron from the less electronegative G to the more electronegative $MoS_2$.

To simulate the 1T/2H hybrid $MoS_2$ contacting with G, we built a supercell model with the 1T/2H phase ratio of 1:1 which is shown in Fig. 3c. Herein the modeled 1T/2H phase boundary resembles the α phase boundary of $MoS_2$ observed in experiment [11]. It can be seen that the positions and atomic bonds of S and Mo atoms at the 1T/2H phase boundary strongly deviate from those of pure 1T and 2H phases, implying strong strain fields in the proximity of the boundary. Surprisingly, charge transfer analysis shows that the boundary part makes a larger contribution than the inner parts in the 1T and 2H phases, reflected by the much larger differential charge density at the boundary in Fig. 3c. The underlying reason for this promoted charge transfer could be due to the modified orbital hybridization and bond distortion of the boundary S/Mo atoms. Our results suggest that the phase boundary part, with its charge transfer ability has been overlooked before, may play a significant role in the charge storage and account for the superb supercapacitive performance shown below.

### *3.2 Electrochemical performance of the $MoS_2$/G electrode*

To evaluate the electrochemical performance of hetero-phase $MoS_2$/G nanohybrids as active supercapacitor electrodes, we performed cyclic voltammetry (CV) measurements in a three-electrode configuration. A platinum foil and Ag/AgCl electrode were used as the counter electrode and the reference electrode, respectively, with an aqueous electrolyte (6 M KOH).

The electrochemical performance of the nanohybrid $MoS_2$/G annealed at different temperatures is shown in Fig. 4a. The CV curves reveal enhanced current densities with increasing temperatures up to 150 °C, implying an increasing amount of phase boundaries due to the growth and more nucleation of these line defects under higher temperature. With further increasing temperature to 200 °C, the electrochemical performance is degraded due to the presence of the dominating 2H phase. To discern the electrochemical contribution from graphene and $MoS_2$ towards the nanohybrid film, we have estimated the specific capacitance of graphene and $MoS_2$ by disregarding any synergetic effect between them. The specific capacitance of $MoS_2$ ($C_{MoS_2}$) was calculated by subtracting the voltammetric catholic charge from the value obtained for graphene in the same window potential through the equation $C_{MoS_2} = (Q_{MoS2/G} – Q_G)/(\Delta V_{MoS2})$. Here, $Q_{MoS2/G}$ and $Q_G$ represent the voltammetric charges of the $MoS_2$/G and graphene electrodes, respectively. By calculating the specific capacitance of graphene from Fig. S4 (supporting information), the capacitance value arising from $MoS_2$ was therefore estimated. Fig. 4b shows the capacitance values for graphene, $MoS_2$ and $MoS_2$/G nanohybrid film. One can see that the capacitance value of graphene almost keep constant in the low temperature range from 50 °C to 200 °C. This is attributed to the promoted conductivity resulting from the reduction of the benzene ring at elevated temperatures. At the temperature of 200 °C, a maximum capacitance value of 133 F $g^{-1}$ can be achieved, which is comparable to those of previously reported graphene or graphene/CNT hybrid film [21].

The electrochemical contribution of $MoS_2$ was supposed to degrade due to the phase transition from 1T to 2H phase at elevated annealing temperatures, as the conductance of 2H-$MoS_2$ is $10^7$ times lower than that of 1T phase [4]. Therefore a higher amount of 1T phase is desired for achieving higher capacitance. We however find that the capacitive performance arising from $MoS_2$ greatly increases from 105 F $g^{-1}$ to 438 F $g^{-1}$ with the increase of the temperature from 50 °C to 150 °C, associated with a reduction of the 1T-$MoS_2$ content from 50% to 34%. Especially, the maximum value of 438 F $g^{-1}$ achieved is almost 4 times as high as that was reported for a $MoS_2$ thin film with 70% 1T-$MoS_2$ phase [4]. These results clearly suggest that the enhanced electrochemical performance of $MoS_2$/G observed in this work is not solely determined by the high concentration of 1T phase.

The enhanced capacitive value achieved in the sample annealed at 150 $^oC$ may well be a result of the optimized hybridization state of 1T/2H phase and the high charge transfer efficiency in the 1T/2H phase boundary as revealed by the first-principles calculation shown in Fig. 3c. Previous study showed that the phase transition from 2H to 1T phase initiates with the formation of the nucleation of 1T phase and then the migration of the 1T-2H phase boundaries with increasing temperature [11]. Similar scenario may exist for the transition from 1T to 2H phase which has a lower transition barrier than the 2H to 1T phase transition. With increasing the temperature from 50 $^oC$ to 200 $^oC$, the 1T phase gradually transforms to the more stable semiconducting 2H phase. Owing to the complex chemical environment and residual stress inherent in the 2D sample, the seeding process and nucleation of the 2H phase in the prepared sample (at 50 $^oC$ or even lower temperature) tend to be random and inhomogeneous. Therefore, with increasing the temperature, the sluggish phase transition occurs through the gradual nucleation and growth of the 1T-2H phase boundaries. At 150 $^oC$ the content of the 1T-2H phase boundaries reach the highest content and gradually decrease after 200 $^oC$ due to the predominant formation of 2H phase which is consistent with Ref. 9. Our predicted higher charge transfer efficiency is also consistent with a recent work which found that the resistance of the 1T/2H nanohybrid $MoS_2$ can be lowered through optimizing the in-plane 1T/2H phase hybridization state [22]. The increasing population of the phase boundaries is also consistent with the promoted current density with increasing temperature as measured in Fig. 4a, with bearing in mind that the newly formed 2H phase is semiconducting and less conducting than the initial 1T phase.

The CV curves as a function of scan rate for $MoS_2$/G annealed at 150 $^oC$ is shown in Fig. 4c. A serial rectangular-shape-like CV curves together with symmetric charge-discharge profiles (supporting information Fig. S5) were observed, reflecting a dominant capacitive contribution from electric double-layer capacitors (EDLCs). Such double-layer capacitive performance clearly shows that the ionic adsorption rather than redox reaction occurs in the $MoS_2$/G nanohybrid film [23]. The corresponding volumetric and gravimetric capacitances are summarized in Fig. 4d. It shows that the $MoS_2$/G electrode is capable of yielding a high volumetric capacitance of 587 F $cm^{-3}$ (231 F $g^{-1}$) at the scan rate of 5 mV $s^{-1}$. Moreover, at a slower scan rate of 1 mV $s^{-1}$, a much higher volumetric capacitance of 685 F $cm^{-3}$ is achieved, which is much higher than that of 200 to 350 F $cm^{-3}$ reported in the re-stacked graphene sheets [24-26]. Such enhanced volumetric capacitances are important for compact and portable energy storage systems where the demand for limited space is a key parameter in consideration [27]. For the filtrated $MoS_2$ films, a volumetric value as high as 400-650 $cm^{-3}$ has been reported [4]. This volumetric capacitance is comparable to our result. We however note that the corresponding gravimetric capacitance of ~100 F $g^{-1}$ (Table 1) [4] is much lower than that of 272 F $g^{-1}$ achieved in the present work. Additionally, a competitive value of 416 F $g^{-1}$, has been reported for the $MoS_2$/G 3D structured matrix [30]. However, the foam-type electrode materials possess an apparently low packing density (< 0.8 g $cm^{-3}$ in most cases), and therefore there exists large empty space within the electrode, which could greatly lower the volumetric capacitance value. In the present work, we therefore clearly demonstrate that the $MoS_2$/G nanohybrid film delivers simultaneously both extra high gravimetric and volumetric energy storage capacitance in this material.

*3.3 Electrochemical performance of $Ni(OH)_2$/G//$MoS_2$/G asymmetric supercapacitor*

An asymmetric supercapacitor (ASC) device was assembled based on the synthesized $MoS_2$/G nanohybrid film. Before assembling the asymmetric full cell device, we firstly synthesized a $Ni(OH)_2$/graphene-based bendable positive electrode ($Ni(OH)_2$/G) via the filtration assisted layer by layer self-assembly route (supporting information Fig. S6). It shows a maximum capacitance of 448 F $g^{-1}$ at a scan rate of 1 mV $s^{-1}$ and a high value of 211

F $g^{-1}$ can be retained at a high scan rate of 100 mV $s^{-1}$. An asymmetric full cell device ($Ni(OH)_2$/G//$MoS_2$/G) was fabricated by using $Ni(OH)_2$/G as the positive electrode and $MoS_2$/G as the negative electrode (Fig. 5). The gel electrolyte of polyvinyl alcohol (PVA)/KOH was used in this study for the all-solid-state device, which shows a good flexibility, an appropriate working temperature with an improved safety.

Fig. 6a shows the CV curves collected at different voltage windows for the $Ni(OH)_2$/G// $MoS_2$/G ASC at the scan rate of 10 mV $s^{-1}$ in PVA/KOH electrolyte. Indeed, the stable electrochemical windows of the ASC can be extended to 1.6 V, which is almost twice as that of conventional carbon-based supercapacitors in aqueous electrolytes (~1 V) [21]. Obviously, the CV response is dependent on the potential, which is different from the potential-independent current response occurring in ECs arising from a nonfaradaic process [31]. The CV curves of the ASC device measured at various scan rates with voltage windows ranging from 0 to 1.6 V are shown in Fig. S7a (supporting information), which indicates the charge storage contributions from both electric double-layer capacitance and pseudocapacitance. Fig. 6b shows the calculated volumetric capacitance and gravimetric capacitance. The solid-state ASC device achieved the maximum volumetric and specific capacitances of 170 F $cm^{-3}$ and 74 F $g^{-1}$ at 5 mV $s^{-1}$, respectively, which are apparently higher than those values recently reported for $Ni(OH)_2$/G//CNT/G (58.5 F $cm^{-3}$/44.8 F $g^{-1}$) [21]. Moreover, the solid-state ASC device shows a good rate capacitance, with 83% of volumetric capacitance being retained when the scan rate increases from 5 to 100 mV $s^{-1}$. The superior performance of the solid-state ASC device was further confirmed by galvanostatic charge/discharge measurements. As shown in Fig. 6c, the charging and discharge curves of the solid-state ASC device are reasonably symmetric, with a good linear relation of discharge/charge voltage versus time. This reflects an ideal capacitive characteristic and rapid charge/discharge property of the ASC device. Fig. 6d compares the specific power density ($P$) and energy density ($W$) of the as-fabricated ASC devices with some of those previously reported nickel-based ones [32]. In this work, the $Ni(OH)_2/G//MoS_2/G$ ASC device with a cell voltage of 1.6 V delivers an energy density of 46.3 mWh $cm^{-3}$ at the power density of 93.6 mW $cm^{-3}$, and it can retain 24.3 mWh $cm^{-3}$ at the power density of 3013 mW $cm^{-3}$. It is also worthy to note that the maximum volumetric energy density of 46.3 mWh $cm^{-3}$ is substantially higher than those reported for graphene-based or $Ni(OH)_2$-based ASCs, including for example a highly porous graphene (19.1 mWh $cm^{-3}$) [25], graphene-based micro-supercapacitors (MPG-MSCs 11.8 mWh $cm^{-3}$) [8], $Ni(OH)_2$/graphene//graphene (4.6 mWh $cm^{-3}$) [31], $Ni(OH)_2$/graphene//CNT/graphene (5 mWh $cm^{-3}$) [21], $MoS_2/Ni(OH)_2$ (2.4 mWh $cm^{-3}$) [33], and even the commercial 4V/500 µAh Li thin-film battery [34].

A long durable and stable cycling life is highly important for energy storage [35]. In this work, the long-term cycling performance of the ASC devices was evaluated in the voltage window of 0-1.6 V at the discharge current density of 4.6 A $cm^{-3}$ for 4000 cycles (Fig. 6e). There is no stumbling in capacitance over the long-term cycling. An increase by 30% of the initial capacitance was observed after the first 600 cycles, and a capacitance retention of 120% is obtained after 4000 cycles. The charge-discharge curves of the electrode at the $1^{st}$, $600^{th}$, and $4000^{th}$ cycles are shown in the inset of Fig. 6e. The discharging time lasts longer at $600^{th}$, and $4000^{th}$ cycles compared to that at the $1^{st}$ cycle, which is consistent with the capacitance retention profile mentioned above. The enhancement rather than degradation of the cycling stability is ascribed to the activation of oxides and dense films which facilitates the intercalation of trapped ions [21]. Furthermore, to verify the mechanical flexibility and robustness of the full cell, the device with an electrode area of 1 cm×1 cm was deformed into different bending states with bending angle of $90^o$, $120^o$, and $180^o$ (supporting information Fig. S7b). The corresponding electrochemical responses were recorded, and the comparisons are shown in Fig. 6f. There is no obvious change in the CV curves observed when the cell was bent to $180^o$, revealing the excellent mechanical flexibility, high robustness, and structural integrity.

## 4. Conclusion

In summary, we have demonstrated a new mechanism achieving high energy/charge storing efficiency via modulating the phase boundary. By intentionally engineering in-plane 1T/2H phase hybridization in the layered $MoS_2$, outstanding gravimetric and volumetric capacitances were derived. Our first-principles calculations suggest that such superb performance was ascribed the high charge transfer effeciency in the 1T/2H hetero-phase boundary and the synergestic effect existing in $MoS_2$/G. The full cell, consisting of an asymmetric supercapacitor with the $MoS_2$/G electrode as the negative electrode and $Ni(OH)_2$/G as the positive electrode, delivers a remarkable electrochemical performance of a high volumetric capacitance of 170 F $cm^{-3}$ and a high volumetric energy and power density (46.3 mWh $cm^{-3}$ and 3.013 $Wcm^{-3}$). Understanding and quantifying the interplay between the evolution of the phase boundary and the energy/charge storage performance provide a major challenge in such phase change materials. Our work highlights the importance of the phase engineering and critical role in the hetero-phase boundary in multi-component nanohybrids for energy storage.

## Conflicts of interest

There are no conflicts of interest to declare in this work.

## Acknowledgements

The authors thank the financial support provided by MOE, Singapore Ministry of Education (Tier 2, MOE2012-T2-2-102), for research conducted at the National University of Singapore. This work was supported by MOE under AcRF Tier 2 (ARC 26/13, No. MOE2013-T2-1-034; ARC 19/15, No. MOE2014-T2-2-093; MOE2015-T2-2-057) and AcRF Tier 1 (RG5/13), and NTU under Start-Up Grant (M4081296.070.500000) in Singapore. We thank Prof. Hua Zhang for useful discussions and suggestions. The authors gratefully acknowledge the use of computing resources at the A*STAR Computational Resource Centre, Singapore

**Table 1 Comparison of gravimetric and volumetric capacitances for $MoS_2$/G with some of those reported in previous literatures**

| Materials | Form | Electrolyte | C*m* (F g$^{-1}$) | C*v* (F cm$^{-3}$) | ***Ref.*** |
|---|---|---|---|---|---|
| $MoS_2$/G | Film | 6 M KOH | **272** | **685** | this work |
| $MoS_2$ | Film | 1 M $H_2SO_4$/ $Li_2SO_4$/ $Na_2SO_4$/ $K_2SO_4$ | ~100 | 400-700 | 4 |
| Graphene | Film | 1 M $H_2SO_4$ | 203 | 255 | 26 |
| $MoS_2$/G | Film | 1 M $Na_2SO_4$ | 11 | 220 | 3-28 |
| $MoS_2$/G | Powders | 1 M $HClO_4$ | 128-265 | - | 23 |
| $MoS_2$/G | Powders | 1 M $Na_2SO_4$ | 243 | - | 5-29 |
| $MoS_2$/G | Foam | 1 M NaCl | 416 | <332 | 28-30 |

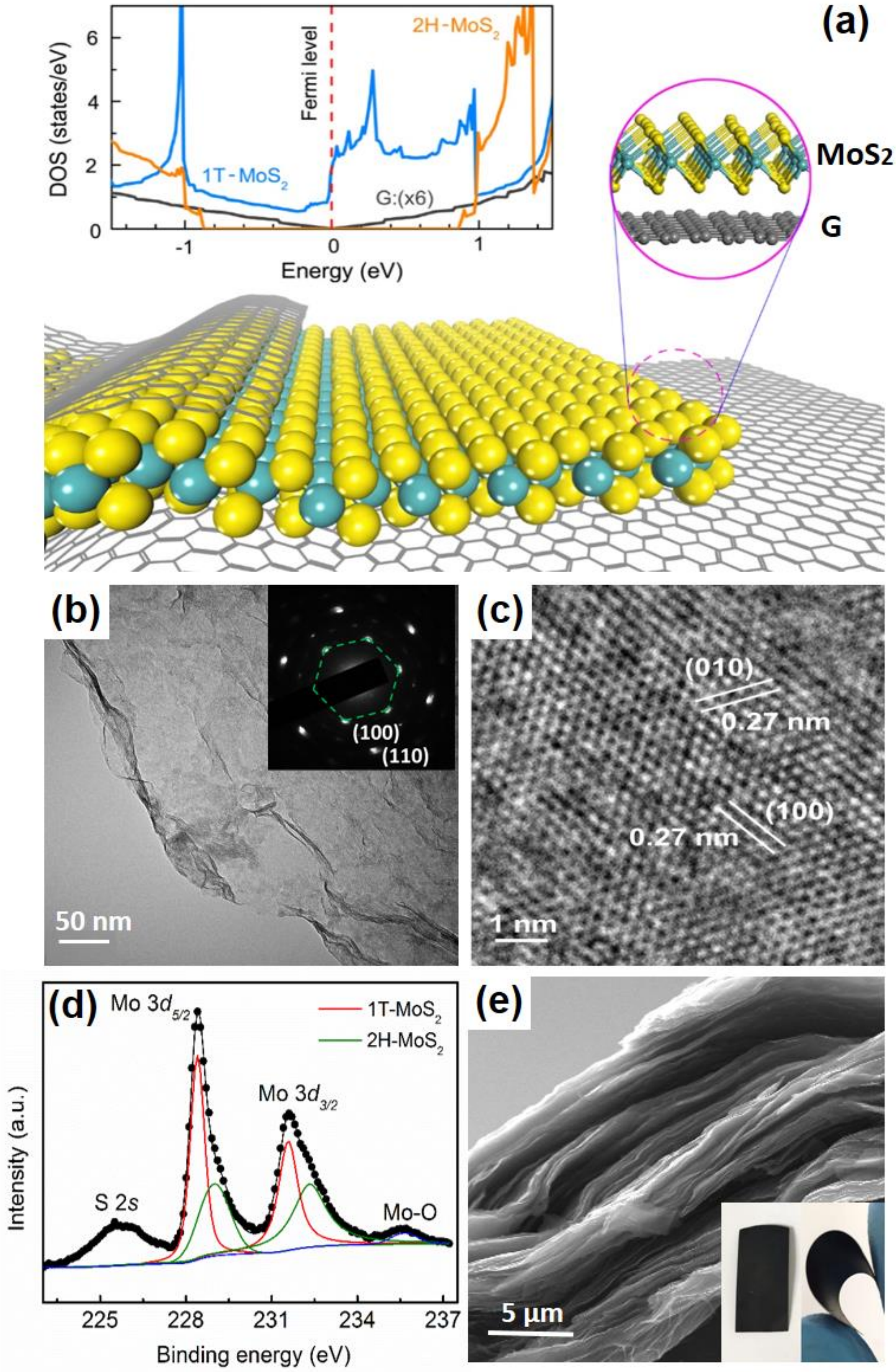

**Fig. 1** (a) Schematic atomic model of the $MoS_2$/G hybrid. Top right inset: Interface formed between G and $MoS_2$. Top left inset: Comparison of the electronic DOS of graphene, 2H- and 1T-$MoS_2$ phases. Note an enlargement of six for the DOS of G. (b) TEM and (c) HRTEM image of $MoS_2$ nanosheets. Inset of b) shows a selected area electron diffraction (SAED) pattern for $MoS_2$ nanosheets. (d) XPS spectra of Mo 3*d* of the as-exfoliated $MoS_2$ nanosheets. (e) SEM images of the side view of the filtrated $MoS_2$/G film, inset optical image showing its flexibility.

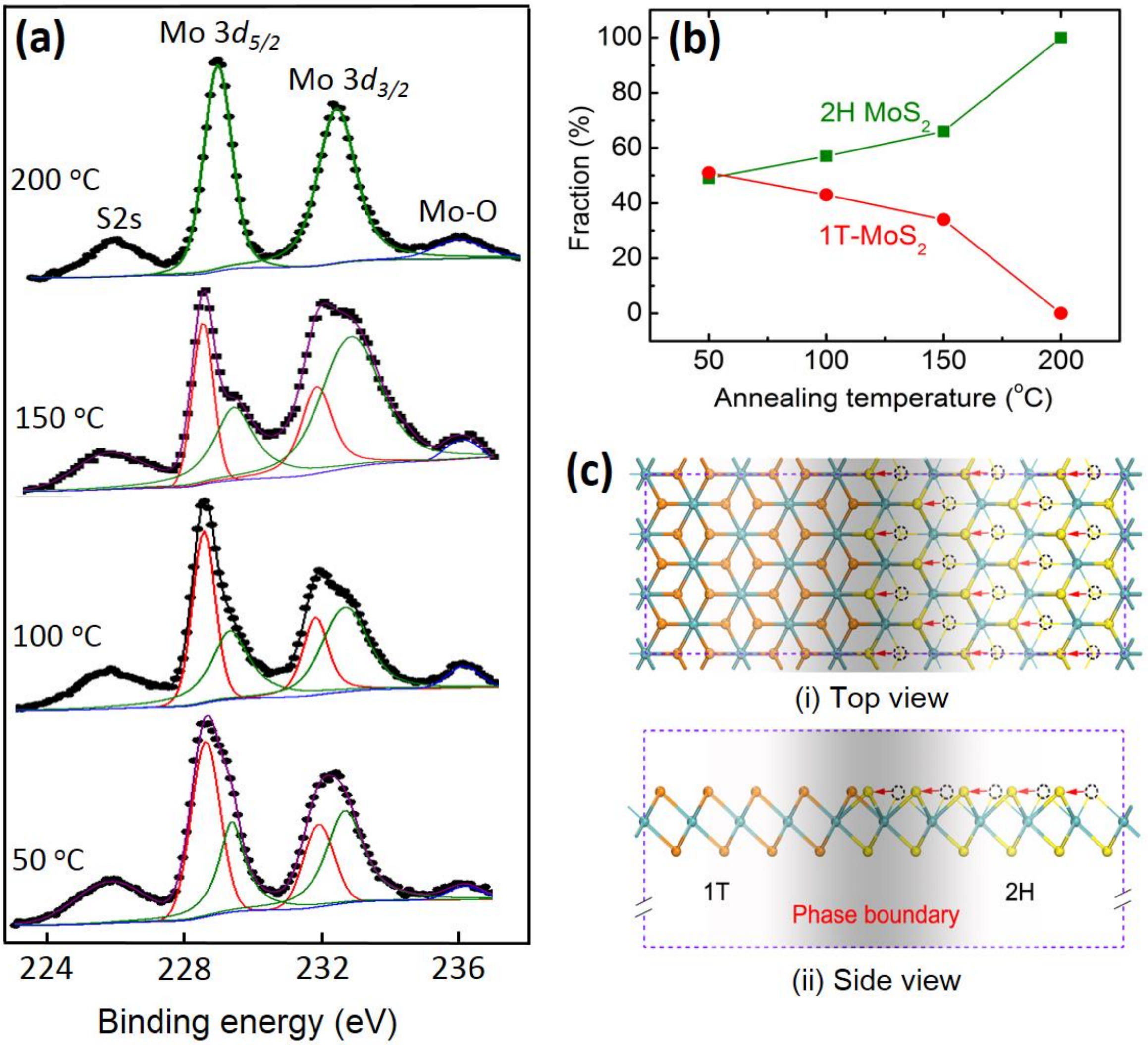

**Fig. 2** (a) XPS spectra showing Mo 3*d* peak regions for samples annealed at various temperatures. After Shirley background subtraction, the Mo 3*d* peaks were deconvoluted to show the contributions of the 2H and 1T phases, represented by green and red plots, respectively. (b) Extracted relative fractions of 2H and 1T components as a function of annealing temperatures are described in green and red dots respectively. (c) Schematic atomic models for a transition from 1T to 2H phase and the formation of the 1T/2H phase boundary (shadowed area). The transition involves the gliding of the top S atoms of the 1T phase along the armchair direction (shown by the red arrows) with leaving holes (dashed circles) in the hollow center of the hexagons, The Mo atoms, S atoms in the 1T phase, S atoms in the 2H phase are represented by blue, purple, and yellow spheres, respectively.

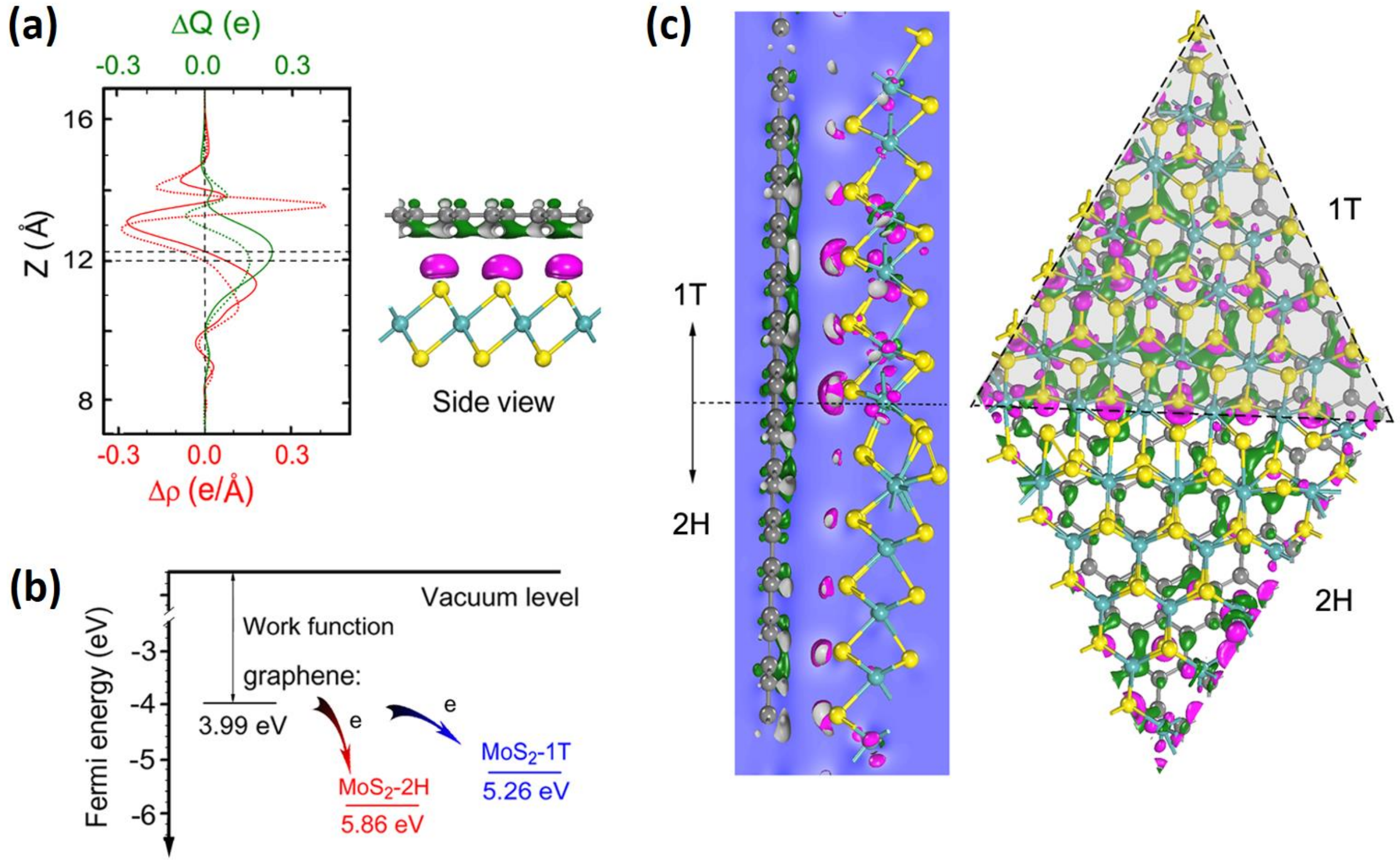

**Fig. 3** (a) The left panel shows the charge transfer across 1T-$MoS_2$/G interface. The green (red) color represents the loss (accumulation) of electrons. The right panel shows the plane-averaged DCD $\Delta\rho(z)$ (red) and $\Delta Q(z)$ (green) curves along the out-of-plane (z) direction for the 1T-$MoS_2$/G (solid lines) and 2H-$MoS_2$/G (dashed lines). The horizontal black dashed line defines the total amount of transferred electrons from graphene to the $MoS_2$ per current supercell. (b) Alignment of the work function of G (3.99 eV), 2H-$MoS_2$ (5.86 eV) and 1T-$MoS_2$ (5.26 eV) calculated by HSE06 method. (c) Side and top views of the DCD which indicates much larger charge transfer at the phase boundary of the 1T/2H heterostructure compared with inner parts of 1T and 2H phases.

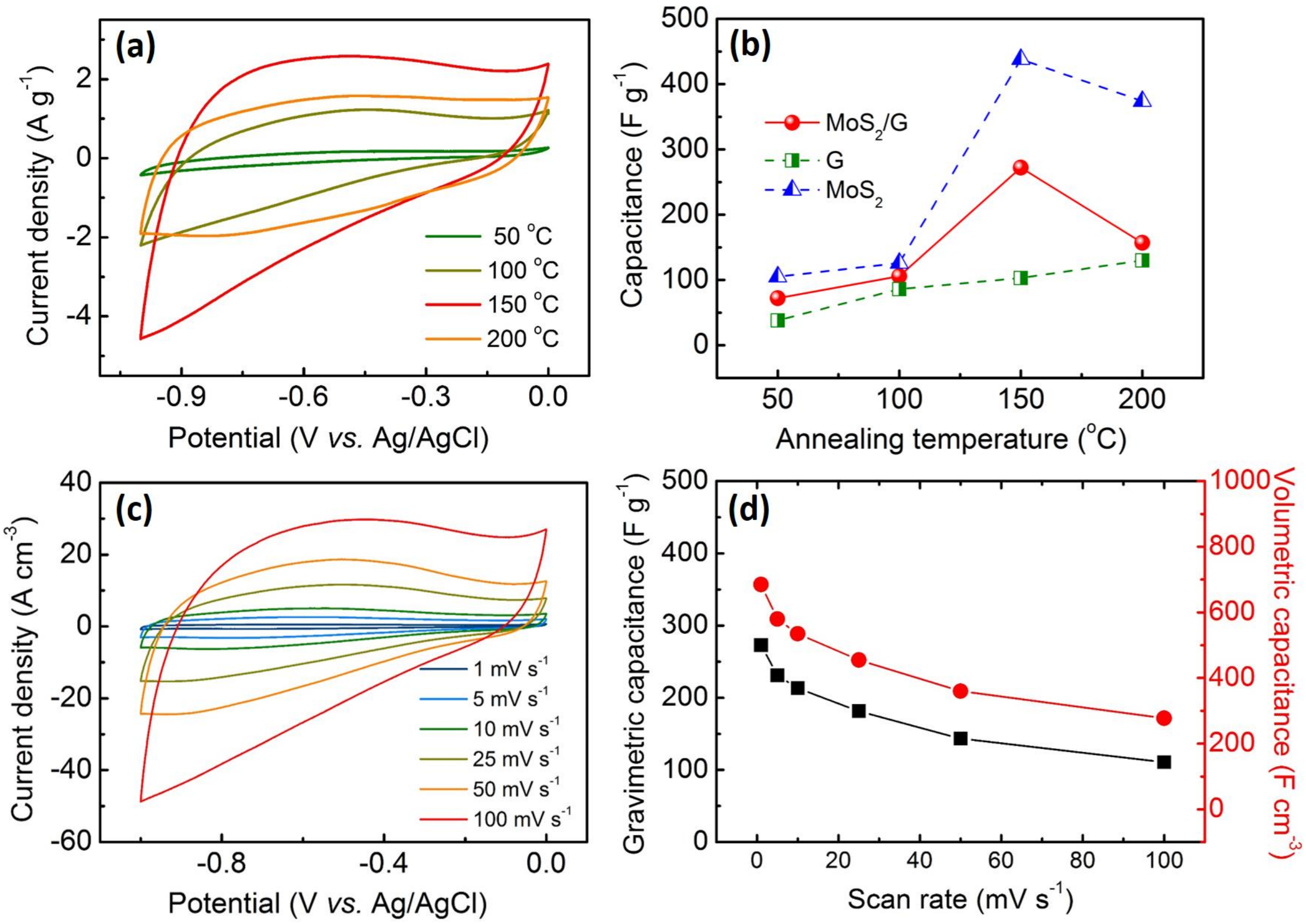

**Fig. 4** (a) Comparison of electrochemical performance of the nanohybrid $MoS_2$/G film annealed at different temperatures. The testing devices are assembled in a three-electrode configuration at the scan rate of 10 mV $s^{-1}$. (b) The capacitance of graphene, $MoS_2$/G, $MoS_2$ as a function of annealing temperatures. (c) CV curves of $MoS_2$/G film measured at different scan rates. (d) Variation of gravimetric and volumetric capacitance of $MoS_2$/G electrode with scan rate.

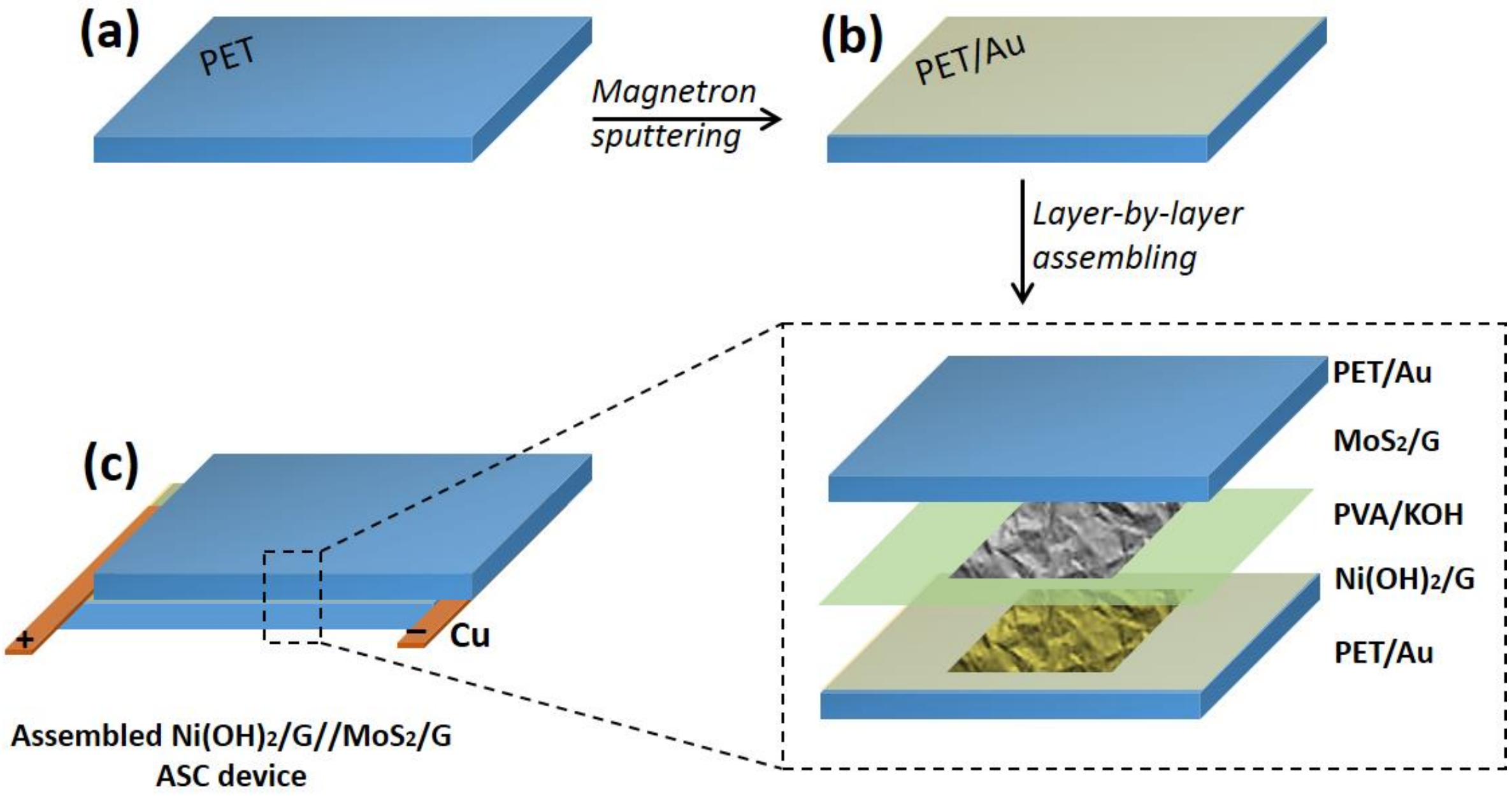

**Fig. 5** Schematic illustration for the fabrication of the $Ni(OH)_2$/G//$MoS_2$/G ASC device. (a) A rectangular flexible PET substrate is used. (b) Growing Au film on PET by magnetron sputtering. (c) Assembly of the sandwiched structure between two pieces of PETs and the detailed cross section including active materials and separator are shown in the right.

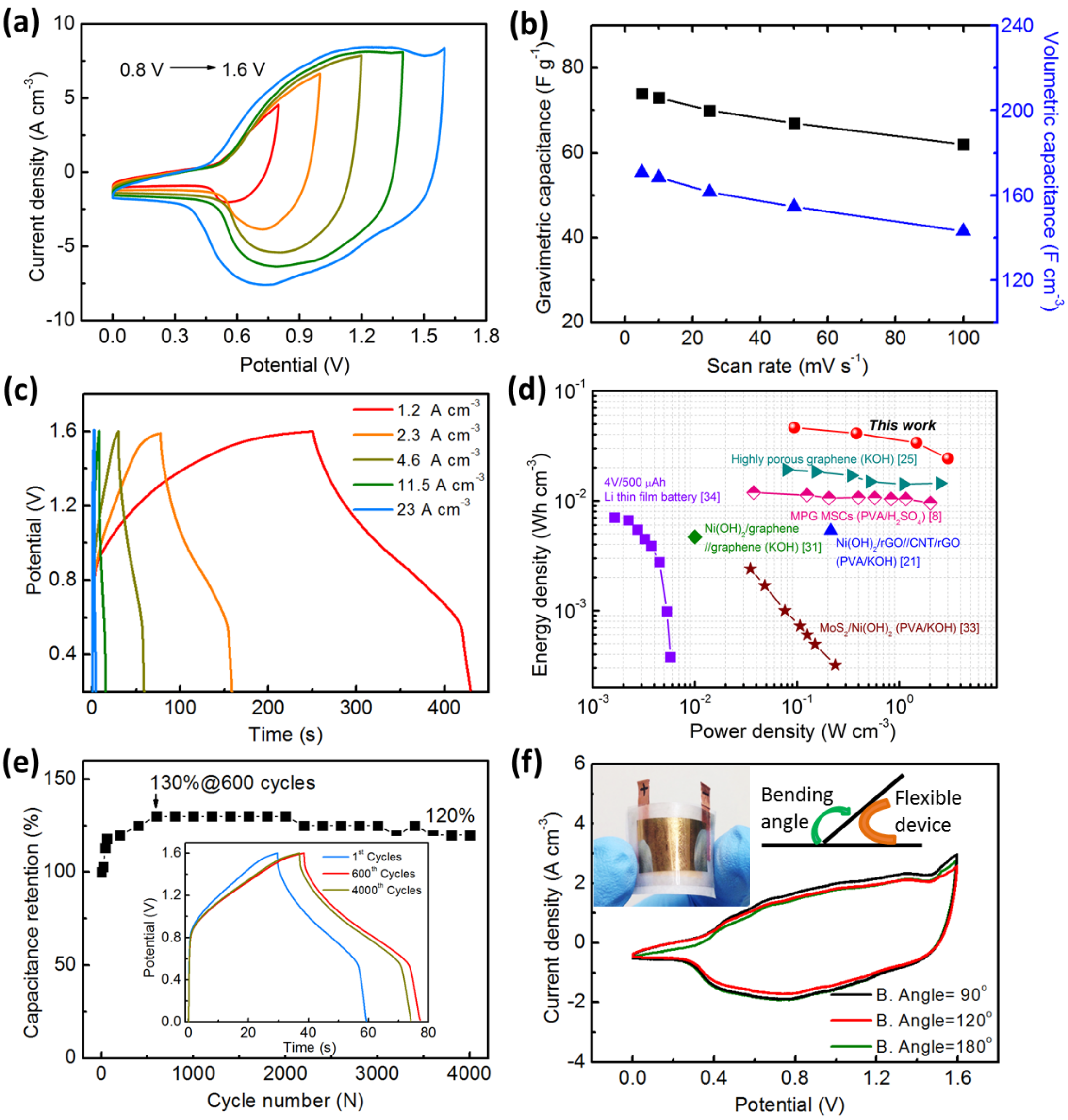

**Fig. 6** Electrochemical performance of the all-solid-state $Ni(OH)_2$/G//$MoS_2$/G ASC device in PVA/KOH. (a) CV curves of the ASC device collected in different scan voltage windows. (b) Variation of gravimetric and volumetric capacitance with scan rate. (c) Galvanostatic charge/discharge curves collected at different current densities for the ASC device operated within a voltage window ranging from 0 to 1.6 V. (d) Ragone plots of the $Ni(OH)_2$/G//$MoS_2$/G ASC device. The volumetric energy and power densities of several commercial energy storage systems and reported devices are plotted and used as references. (e) Cyclability of the $Ni(OH)_2$/G//$MoS_2$/G ASC device in PVA/KOH electrolyte. Inset shows the galvanostatic charge/discharge curves variation at different stages of cycling test. (f) The CV curves of the solid-state ASC collected at a scan rate of 10 mV $s^{-1}$ in different bending states, inset showing the optical photograph of the fabricated bendable ASC device and the related definition of the bending angles (B. Angle).